\documentclass{article}
\pdfoutput=1

\usepackage{amsmath,amssymb,array,multirow,tikz,tikz-cd,rotating}

\usepackage{amsthm}
\theoremstyle{definition}
\newtheorem{definition}{Definition}

\theoremstyle{theorem}
\newtheorem{proposition}{Proposition}
\newtheorem{lemma}{Lemma}
\newtheorem{theorem}{Theorem}

\theoremstyle{definition}

\newtheorem{fact}{Fact}

\usetikzlibrary{arrows,automata}

\usepackage{hyperref}

\newcommand{\PFA}{\textsf{A}}
\newcommand{\PFB}{\textsf{B}}
\newcommand{\PFC}{\textsf{C}}
\newcommand{\PFE}{\textsf{E}}
\newcommand{\PFF}{\textsf{F}}
\newcommand{\PFG}{\textsf{G}}
\newcommand{\PFH}{\textsf{H}}
\newcommand{\PFI}{\textsf{I}}
\newcommand{\PFL}{\textsf{L}}
\newcommand{\PFM}{\textsf{M}}
\newcommand{\PFN}{\textsf{N}}
\newcommand{\PFO}{\textsf{O}}
\newcommand{\PFQone}{\textsf{Q1}}
\newcommand{\PFQtwo}{\textsf{Q2}}
\newcommand{\PFQthree}{\textsf{Q3}}
\newcommand{\PFQfour}{\textsf{Q4}}
\newcommand{\PFPS}{\textsf{PS}}
\newcommand{\PFHC}{\textsf{HC}}
\newcommand{\PFST}{\textsf{ST}}
\newcommand{\PFAD}{\textsf{AD}}
\newcommand{\PFLE}{\textsf{LE}}
\newcommand{\PFSR}{\textsf{SR}}
\newcommand{\PFAW}{\textsf{AW}}
\newcommand{\PFPI}{\textsf{PI}}
\newcommand{\PFOT}{\textsf{OT}}
\newcommand{\PFAP}{\textsf{AP}}
\newcommand{\PFTS}{\textsf{TS}}
\newcommand{\PFTI}{\textsf{TI}}

\newcommand{\sixteenPF}{\mbox{$16\mathbb{PF}$}}
\newcommand{\PsychEvalPF}{\mathbb{PEPF}}

\newcommand{\PPPts}{\textrm{PPP}}
\newcommand{\PPPps}{\mathcal{PPP}}
\newcommand{\PPP}{\mathbb{PF}}

\newcommand{\mbbb}{-!!!}
\newcommand{\mbb}{-!!}
\newcommand{\mb}{-!}
\newcommand{\m}{-}
\newcommand{\n}{0}
\newcommand{\p}{+}
\newcommand{\pb}{+!}
\newcommand{\pbb}{+!!}
\newcommand{\pbbb}{+!!!}
\newcommand{\pmlb}{\pm_{!}}
\newcommand{\pmub}{\pm^{!}}

\newcommand{\F}[2]{\mbox{$\textsf{#1}#2$}}

\newcommand{\signatures}{\mathbb{S}}
\newcommand{\factors}{\mathbb{F}}

\newcommand{\SPPts}{\textrm{SPP}}
\newcommand{\SPPps}{\mathcal{SPP}}

\newcommand{\atoms}{\mathbb{A}}
\newcommand{\LPL}{\textrm{LPL}}
\newcommand{\LPLbis}{\mathcal{LPL}}

\newcommand{\rightG}[1]{{#1^{\triangleright}}}
\newcommand{\leftG}[1]{{#1^{\triangleleft}}}
\newcommand{\rightleftG}[1]{{#1^{\triangleright\triangleleft}}}
\newcommand{\leftrightG}[1]{{#1^{\triangleleft\triangleright}}}
\newcommand{\rightI}{\mathrm{f}}
\newcommand{\leftI}{\mathrm{p}}

\newcolumntype{C}{>{\centering\arraybackslash}p{0.26\textwidth}}

\begin{document}
\title{A Galois-Connection between Cattell's and Szondi's Personality Profiles}
\author{Simon Kramer\\[\jot]
		\texttt{simon.kramer@a3.epfl.ch}}
\maketitle
\begin{abstract}
We propose a computable Galois-connection between,  
	on the one hand,  
		\emph{Cattell's 16-Personality-Factor (16PF) Profiles,} 
			one of the most comprehensive and widely-used personality measures for non-psychiatric populations and 
		their containing \emph{PsychEval Personality Profiles (PPPs)} for psychiatric populations, and, 
	on the other hand, 
		\emph{Szondi's personality profiles (SPPs),}  
		a less well-known but, as we show, finer personality measure for 
			psychiatric as well as non-psychiatric populations 
				(conceived as a unification of the depth psychology of S.\  Freud, C.G.\ Jung, and A.\ Adler).
The practical significance of our result is that 
	our Galois-connection provides a pair of computable, 
		interpreting translations between the two personality spaces of 
				PPPs (containing the 16PFs) and SPPs: 
			one \emph{concrete} from PPP-space to SPP-space (because SPPs are finer than PPPs) and 
			one \emph{abstract} from SPP-space to PPP-space (because PPPs are coarser than SPPs).
Thus Cattell's and Szondi's personality-test results are 
	mutually interpretable and inter-translatable, 
		even automatically by computers.
	
	\smallskip
	
\end{abstract}

\section{Introduction}
According to \cite[Page~3]{16PFinHB},
	most studies have found 
		Cattell's comprehensive 16-Personality-Factor (16PF) Profiles \cite{16PF_Questionnaire}  
		``to be among the top five most commonly used normal-range instruments in both research and practice'' 
			with culturally adapted translations into over 35 languages world-wide.
Further,
	``[t]he 16PF traits also appear in the PsychEval Personality Questionnaire \cite{PsychEval_Questionnaire}, 
		a comprehensive instrument which includes both normal and abnormal personality dimensions.''
Note that according to \cite[Page~4]{16PFinHB},  
	``[i]nstead of being developed to measure preconceived dimensions of interest to a particular author, 
		the instrument was developed from the unique perspective of a scientific quest to try to discover 
			the basic structural elements of personality.''
Notwithstanding, and further exemplifying our general methodology introduced in \cite{arXiv:1403.2000v1}, 
	we propose in the present paper a computable Galois-connection \cite{DaveyPriestley} between 
		\emph{PsychEval Personality Profiles (PPPs),} which contain the 16PFs, and 
		\emph{Szondi's Personality Profiles (SPPs)} \cite{Szondi:ETD:Band1}, 
			a less well-known but, as we show, finer personality measure for 
				psychiatric as well as non-psychiatric populations, and 
					conceived as a unification \cite{Szondi:IchAnalyse} of the depth psychology of 
						S.\ Freud, C.G.\ Jung, and A.\ Adler.
This paper being a further illustration of our general methodology introduced in \cite{arXiv:1403.2000v1},
our presentation here thus closely follows the one in \cite{arXiv:1403.2000v1}, even in wording.
The generality of our mathematical methodology may be obvious to the (order-theoretic) mathematician, but 
	may well not be so to the general psychologist.

Just like \cite{arXiv:1403.2000v1}, 
	our present result 
		is a contribution to \emph{mathematical psychology} in the area of 
			\emph{personality assessment.}
It is also meant as a contribution towards 
	practicing psychological research with the methods of 
		the exact sciences, for
			obvious ethical reasons.
The practical significance of our result is that 
		our Galois-connection provides a pair of computable, 
			interpreting translations between the two personality spaces of PPPs and SPPs 
				(and thus hopefully also between their respective academic and non-academic communities): 
				one \emph{concrete} translation from PPP-space to SPP-space (because SPPs are finer than PPPs) and
				one \emph{abstract} translation from SPP-space to PPP-space (because PPPs are coarser than SPPs).
Thus Cattell's and Szondi's personality-test results are  
	mutually interpretable and inter-translatable, 
		even automatically by computers.
The only restriction to this mutuality is 
	the subjective interpretation of the faithfulness of these translations.
In our interpretation,
	we intentionally restrict the translation from SPP-space to PPP-space, and only that one, 
		in order to preserve (our perception of) its faithfulness. 
More precisely,
	we choose to map some SPPs to the empty set in PPP-space
		(but every PPP to a non-empty set in SPP-space).
Of course just like in \cite{arXiv:1403.2000v1}, 
	our readers can 
		experiment with their own interpretations, 
			as we explain again in the following paragraph.
		
We stress that 
	our Galois-connection between the spaces of PPPs and SPPs is 
		independent of their respective \emph{test,} which 
			evaluate their testees in terms of 
				\emph{structured result values}---the PPPs and SPPs---in the respective space.
Both tests are preference-based, more precisely, 
	test evaluation is based 
		on choices of preferred questions in the case of the PsychEval-test \cite{PsychEval_Questionnaire} and 
		on choices of preferred portraits in the case of the Szondi-test \cite{Szondi:ETD:Band1,SzondiTestWebApp}.
Due to the independence of our Galois-connection from these tests,
	their exact nature need not concern us here.
All what we need to be concerned about is the nature of the structured result values that these tests generate.
(Other test forms can generate the same form of result values, e.g.~\cite{Kenmo:Szondi}.)
We also stress 
	that our proposed Galois-connection is 
		what we believe to be an interesting candidate brain child for adoption by the community, but
	that there are other possible candidates, which our readers are empowered to explore themselves.
In fact,
	not only 
		do we propose a candidate Galois-connection between PPP-space and SPP-space, but also 
		do we further illustrate the whole \emph{methodology} introduced in \cite{arXiv:1403.2000v1} for 
			generating such candidates.
All what 
	readers interested in generating such connections themselves need to do is 
		map their own intuition about 
			the meaning of PPPs to a standard interlingua, 
				called \emph{Logical Pivot Language (LPL)} here, and check that 
					their mapping has a single simple property,
						namely the one stated as Fact~\ref{fact:FactsAboutip}.1 about 
							our mapping $\rightI$ in  
								Figure~\ref{figure:MappingsAndMorphisms}.
Their desired Galois-connection is then automatically induced jointly by 
	their chosen mapping and 
	a mapping, called $\leftI$, from SPP-space to LPL that
		we chose in \cite{arXiv:1403.2000v1} once and for all possible Galois-connections of interest.
What is more, and as already mentioned in \cite{arXiv:1403.2000v1} and evidenced here, 
	our methodology is applicable even more generally to the generation of Galois-connections between 
		pairs of result spaces of other personality tests.
SPPs just happen to have a finer structure than 
	other personality-test values that we are aware of, and 
		so are perhaps best suited to play 
			the distinguished role of explanatory semantics for result values of other personality tests.
Of course our readers are still free to choose their own preferred semantic space.
	
An SPP can be conceived as a tuple of eight, 
	so-called \emph{signed factors} whose signatures can in turn 
		take \emph{12 partially ordered} values.
So SPPs live in an eight-dimensional space.
On the other hand,
	a PPP can be conceived as a (16+12=28)-tuple of so-called \emph{personality traits,} which
		can take \emph{10 totally ordered} values.
So PPPs live in an apparently finer, 28-dimensional space.
Nevertheless,
	we are going to show that actually the opposite is true, that is,
		SPPs are finer than PPPs.
In particular, 
	SPPs can account for \emph{ambiguous personality traits} thanks to the partiality of their ordering, whereas 
		PPPs cannot due to the totality of theirs.
Moreover, 
	a lot of Cattell's personality traits turn out to be definable in terms of 
		a combination of Szondi's signed factors, which
			means that 
				a lot of Cattell's personality traits can be understood as 
					(non-atomic/-primitive) \emph{psychological syndromes.}
SPPs being finer than PPPs,
	the translation from SPPs to PPPs must be a projection (and thus surjection) of SPP-space onto PPP-space.  
Another insight gained in the finer referential system of SPPs is that  		
	PPPs are confirmed to be non-orthogonal or not independent as also mentioned in \cite{16PFinHB}.
Of course our readers are still free to disagree on the value of these insights by
	giving a convincing argument for why SPP-space would be an inappropriate semantics for PPP-space.
After all,
	Szondi conceived his theory of human personality as 
		a unifying theory.
We now put forward our own argument for why we believe SPP-space is indeed  
	an appropriate---though surely not the only---semantics for PPP-space.
In Section~\ref{section:Structures},
	we present the defining mathematical structures for each space, and 
in Section~\ref{section:MappingsAndMorphisms},
	the defining mathematical mappings for their translation.
No prior knowledge of either PPPs or SPPs is required to appreciate the results of this paper, but
	the reader might appreciate them even more when comparing them also with those in \cite{arXiv:1403.2000v1}.

\section{The connection}
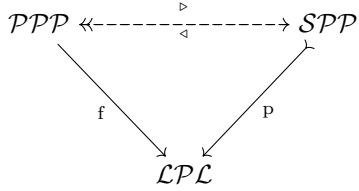
\begin{figure}
	\caption{Mappings between personality spaces and interlingua}
	$$\begin{tikzcd}
		\PPPps \arrow[swap]{ddr}{\rightI} \arrow[yshift=1ex, dashed]{rr}{\rightG{}} & & \SPPps \arrow[tail]{ddl}{\leftI} \arrow[dashed, yshift=-0.5ex, two heads, name=T, below]{ll}{\leftG{}}\\
		&&\\
		& \LPLbis &
	\end{tikzcd}$$
	\label{figure:MappingsAndMorphisms}
\end{figure}

In this section, 
	we present 
		the defining mathematical structures for 
			PPP-space, the interlingua LPL, and SPP-space, as well as
		the defining mathematical mappings for 
			the concrete translation of PPP-space to SPP-space and 
			the abstract translation of SPP-space back to PPP-space, both via LPL, 
				see Figure~\ref{figure:MappingsAndMorphisms}.
				
\subsection{Structures}\label{section:Structures}
In this section, 
	we present 
		the defining mathematical structures for 
			PPP-space, the interlingua LPL, and SPP-space.
We start with defining PPP-space.
\begin{definition}[The PsychEval Personality Profile Space]
Let 
	\begin{itemize}
		\item $\sixteenPF=\{\, \PFA, \PFB, \PFC, \PFE, \PFF, \PFG, \PFH, \PFI, \PFL, \PFM, \PFN, \PFO, \PFQone, \PFQtwo, \PFQthree, \PFQfour\, \}$ be the set of the \emph{16 Personality Factors} (the normal traits), with 
				\PFA\ meaning ``warmth,'' 
				\PFB\ ``reasoning,'' 
				\PFC\ ``emotional stability,'' 
				\PFE\ ``dominance,'' 
				\PFF\ ``liveliness,'' 
				\PFG\ ``rule-consciousness,'' 
				\PFH\ ``social boldness,'' 
				\PFI\ ``sensitivity,'' 
				\PFL\ ``vigilance,'' 
				\PFM\ ``abstractness,'' 
				\PFN\ ``privateness,'' 
				\PFO\ ``apprehension,'' 
				\PFQone\ ``openness to change,'' 
				\PFQtwo\ ``self-reliance,'' 
				\PFQthree\ ``perfectionism,'' and 
				\PFQfour\  ``tension;''
		\item $\PsychEvalPF=\{\, \PFPS, \PFHC, \PFST, \PFAD, \PFLE, \PFSR, \PFAW, \PFPI, \PFOT, \PFAP, \PFTS, \PFTI\, \}$ be the set of the \emph{12 PsychEval abnormal traits,} with 
				\PFPS\ meaning ``psychological inadequacy,'' 
				\PFHC\ ``health concerns,'' 
				\PFST\ ``suicidal thinking,'' 
				\PFAD\ ``anxious depression,'' 
				\PFLE\ ``low energy state,'' 
				\PFSR\ ``self-reproach,'' 
				\PFAW\ ``apathetic withdrawal,'' 
				\PFPI\ ``paranoid ideation,'' 
				\PFOT\ ``obsessional thinking,'' 
				\PFAP\ ``alienation/perceptual distortion,'' 
				\PFTS\ ``thrill seeking,'' and 
				\PFTI\ ``threat immunity;''
		\item $\PPP=\sixteenPF\cup\PsychEvalPF$\,.
	\end{itemize}
Then,
	$$\PPPts=\{\; \begin{array}[t]{@{}l@{}}
				(\begin{array}[t]{@{}l@{}}
					(\PFA,v_{1}),(\PFB,v_{2}),(\PFC,v_{3}),(\PFE,v_{4}),(\PFF,v_{5}),(\PFG,v_{6}),(\PFH,v_{7}),(\PFI,v_{8}),(\PFL,v_{9}),\\
					(\PFM,v_{10}),(\PFN,v_{11}),(\PFO,v_{12}),(\PFQone,v_{13}),(\PFQtwo,v_{14}),(\PFQthree,v_{15}),(\PFQfour,v_{16}),\\
				 	(\PFPS,v_{17}),(\PFHC,v_{18}),(\PFST,v_{19}),(\PFAD,v_{20}),(\PFLE,v_{21}),(\PFSR,v_{22}),\\					(\PFAW,v_{23}),(\PFPI,v_{24}),(\PFOT,v_{25}),(\PFAP,v_{26}),(\PFTS,v_{27}),(\PFTI,v_{28}))\mid
					\end{array}\\[12.5\jot] 
				 v_{1},\ldots,v_{28}\in\{1,2,3,4,5,6,7,8,9,10\}\;\}
				 \end{array}$$
is the set of PsychEval Personality Profiles (PPPs) \cite{16PF_Questionnaire,PsychEval_Questionnaire}, and
	$$\PPPps=\langle\, 2^{\PPPts},\emptyset,\cap,\cup,\PPPts,\overline{\,\cdot\,},\subseteq\,\rangle$$
defines our \emph{PsychEval Personality Profile Space,} that is,
	the (inclusion-ordered, Boolean) powerset algebra \cite{DaveyPriestley} on \PPPts\ 
		(the set of all subsets of \PPPts).
\end{definition}
\noindent
Note that 
	we do need to define $\PPPps$ as the set of all \emph{subsets} of $\PPPts$ and 
		not simply as the set of all elements of $\PPPts$.
The reason is the aforementioned fact that 
	in the finer referential system of SPP-space (see Definition~\ref{definition:SPP}), 
		PPPs turn out to be non-orthogonal or not independent, and thus 
			a PPP may have to be mapped to a proper set of SPPs (see Table~\ref{table:PPPtoLPL}).
So the proper setting for SPP-space is a set of \emph{subsets} of SPPs, which
	in turn, via the backward translation from SPP-space to $\PPPps$, means that 
		the proper setting for $\PPPps$, as the target of a mapping of subsets, 
			is also a set of subsets.

We continue to define SPP-space.
\begin{definition}[The Szondi Personality Profile Space]\label{definition:SPP}
Let us consider the Hasse-diagram \cite{DaveyPriestley} in Figure~\ref{figure:SzondiSignatures} 
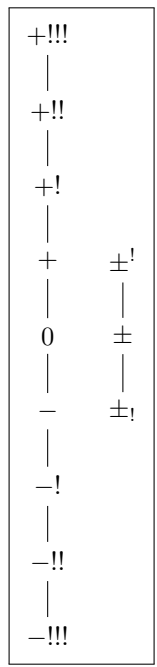
\begin{figure}[t]
\centering
\caption{Hasse-diagram of Szondi's signatures}
\medskip
\fbox{\begin{tikzpicture}
	\node (pbbb) at (0,4) {$+!!!$};
	\node (pbb) at (0,3) {$+!!$};
	\node (pb) at (0,2) {$+!$};
	\node (p) at (0,1) {$+$};
	\node (n) at (0,0) {$0$};
	\node (m) at (0,-1) {$-$};
	\node (mb) at (0,-2) {$-!$};
	\node (mbb) at (0,-3) {$-!!$};
	\node (mbbb) at (0,-4) {$-!!!$};
	\draw (mbbb) -- (mbb) -- (mb) -- (m) -- (n) -- (p) -- (pb) -- (pbb) -- (pbbb);	
 	\node (pmub) at (1,1) {$\pm^{!}$};  
	\node (pm) at (1,0) {$\pm$};
	\node (pmlb) at (1,-1) {$\pm_{!}$};
	\draw (pmlb) -- (pm) -- (pmub);
\end{tikzpicture}}
\label{figure:SzondiSignatures}
\end{figure}
of the partially ordered set of \emph{Szondi's twelve signatures} \cite{Szondi:ETD:Band1} of 
human reactions, which are:
\begin{itemize}
	\item approval: from strong $+!!!$\,, $+!!$\,, and $+!$ to weak $+$\,; 
	\item indifference/neutrality: $0$\,; 
	\item rejection: from weak $-$\,, $-!$\,, and $-!!$ to strong $-!!!$\,; and 
	\item ambivalence: $\pm^{!}$ (approval bias), $\pm$ (no bias), and $\pm_{!}$ (rejection bias).
\end{itemize}
(Szondi calls the exclamation marks in his signatures \emph{quanta.})

Further let us call this set of signatures $\mathbb{S}$, that is,
		$$\signatures=\{\,\mbbb,\mbb,\mb,\m,\n,\p,\pb,\pbb,\pbbb,\pmlb,\pm,\pmub\,\}.$$	
	
Now let us consider \emph{Szondi's eight factors and four vectors} of 
	human personality \cite{Szondi:ETD:Band1} as summarised in Table~\ref{table:SzondiFactors}.
	\begin{table}[t]
		\centering
		\caption{Szondi's factors and vectors}
		\medskip
		{\small
		\begin{tabular}{|c|c||C|C|}
			\hline
			\multirow{2}{12.5ex}{\centering \textbf{Vector}} & \multirow{2}{7.75ex}{\textbf{Factor}} & \multicolumn{2}{c|}{\textbf{Signature}}\\
			\cline{3-4}
			&& $+$ & $-$\\
			\hline
			\hline
			\multirow{2}{12.5ex}{\centering \textsf{S} (Id)} & \textsf{h} (love) & physical love & platonic love\\
			\cline{2-4}
			& \textsf{s} (attitude) & (proactive) activity & (receptive) passivity\\
			\hline
			\multirow{2}{12.75ex}{\centering \textsf{P}\\[-\jot] (Super-Ego)} & \textsf{e} (ethics) & ethical behaviour & unethical behaviour\\
			\cline{2-4}
			& \textsf{hy} (morality) & immoral behaviour & moral behaviour\\
			\hline
			\multirow{2}{12.5ex}{\centering \textsf{Sch} (Ego)} & \textsf{k} (having) & having more & having less\\
			\cline{2-4}
			& \textsf{p} (being) & being more & being less\\
			\hline			
			\multirow{2}{12.5ex}{\centering \textsf{C} (Id)} & \textsf{d} (relations) & unfaithfulness & faithfulness\\
			\cline{2-4}
			& \textsf{m} (bindings) & dependence & independence\\
			\hline		
		\end{tabular}}
		\label{table:SzondiFactors}
	\end{table}
(Their names are of clinical origin and need not concern us here.)
And let us call the set of factors $\factors$, that is, 
	$$\factors=\{\,\F{h}{},\F{s}{},\F{e}{},\F{hy}{},\F{k}{},\F{p}{},\F{d}{},\F{m}{}\,\}.$$

Then,
	$$\SPPts=\{\; \begin{array}[t]{@{}l@{}}
				((\F{h}{,s_{1}}), (\F{s}{,s_{2}}), (\F{e}{,s_{3}}), (\F{hy}{,s_{4}}), 
				 (\F{k}{,s_{5}}), (\F{p}{,s_{6}}), (\F{d}{,s_{7}}), (\F{m}{,s_{8}})) \mid\\ 
				 s_{1},\ldots,s_{8}\in\signatures\;\}
				 \end{array}$$
is the set of Szondi's personality profiles, and
	$$\SPPps=\langle\, 2^{\SPPts},\emptyset,\cap,\cup,\SPPts,\overline{\,\cdot\,},\subseteq\,\rangle$$
defines our \emph{Szondi Personality Profile Space,} that is,
	the (inclusion-ordered, Boolean) powerset algebra \cite{DaveyPriestley} on \SPPts\ 
		(the set of all subsets of \SPPts).
\end{definition}
\noindent
As an example of an SPP,
	consider the \emph{norm profile} for the Szondi-test \cite{Szondi:ETD:Band1}:
		$$((\F{h}{,\p}), (\F{s}{,\p}), (\F{e}{,\m}), (\F{hy}{,\m}), 
				 (\F{k}{,\m}), (\F{p}{,\m}), (\F{d}{,\p}), (\F{m}{,\p}))$$
Spelled out, 
	this norm profile describes the personality of a human being who 
		approves of physical love,
		has a proactive attitude,
		has unethical but moral behaviour,
		wants to have and be less, and 
		is unfaithful and dependent.

We conclude this subsection with the definition of our interlingua LPL.
\begin{definition}[The Logical Pivot Language]
	Let 
		$$\atoms=\{\,\F{h}{s_{1}}, \F{s}{s_{2}}, \F{e}{s_{3}}, \F{hy}{s_{4}}, \F{k}{s_{5}}, \F{p}{s_{6}}, \F{d}{s_{7}}, \F{m}{s_{8}}  \mid  
					s_{1},\ldots,s_{8}\in\signatures\,\}$$
	be our set of atomic logical formulas, and 
		$\LPL(\atoms)$ the classical propositional language over $\atoms$, that is, 
		the set of sentences constructed from the elements in $\atoms$ and
			the classical propositional connectives 
				$\neg$ (negation, pronounced ``not''),
				$\land$ (conjunction, pronounced ``and''),
				$\lor$ (disjunction, pronounced ``or''), etc.
	
	Then,
		$$\LPLbis=\langle\,\LPL(\atoms),\Rightarrow\,\rangle$$
	defines our \emph{logical pivot language,} with
		$\Rightarrow$ being logical consequence.
		
	Logical equivalence $\equiv$ is defined in terms of $\Rightarrow$ such that 
		for every $\phi,\varphi\in\LPL(\atoms)$, 
			$\phi\equiv\varphi$ by definition if and only if 
				$\phi\Rightarrow\varphi$ and $\varphi\Rightarrow\phi$.
\end{definition}

\subsection{Mappings between structures}\label{section:MappingsAndMorphisms}
In this section, 
	we present 
		the defining mathematical mappings for 
			the concrete translation $\rightG{}$ of $\PPPps$ to $\SPPps$ via $\LPLbis$ and 
			the abstract translation $\leftG{}$ of $\SPPps$ back to $\PPPps$ again via $\LPLbis$ by 
				means of the auxiliary mappings $\rightI$ and $\leftI$.
We also prove that the ordered pair $(\,\rightG{},\leftG{}\,)$ is a Galois-connection, as promised.
\begin{definition}[Mappings]\label{definition:MappingsAndMorphisms}
	Let the mapping (total function)
		\begin{itemize}
			\item $\rightI$ be defined in 
				\begin{itemize}
					\item the function space $((\PPP\times\{1,\ldots,10\})\to\LPL(\atoms))$ as in Table~\ref{table:PPPtoLPL},
					\item the function space $(\PPPts\to\LPL(\atoms))$ such that 
					$$\begin{array}{@{}r@{}}
						 \rightI((\begin{array}[t]{@{}l@{}}
					(\PFA,v_{1}),(\PFB,v_{2}),(\PFC,v_{3}),(\PFE,v_{4}),(\PFF,v_{5}),(\PFG,v_{6}),(\PFH,v_{7}),(\PFI,v_{8}),(\PFL,v_{9}),\\
					(\PFM,v_{10}),(\PFN,v_{11}),(\PFO,v_{12}),(\PFQone,v_{13}),(\PFQtwo,v_{14}),(\PFQthree,v_{15}),(\PFQfour,v_{16}),\\
				 	(\PFPS,v_{17}),(\PFHC,v_{18}),(\PFST,v_{19}),(\PFAD,v_{20}),(\PFLE,v_{21}),(\PFSR,v_{22}),\\					(\PFAW,v_{23}),(\PFPI,v_{24}),(\PFOT,v_{25}),(\PFAP,v_{26}),(\PFTS,v_{27}),(\PFTI,v_{28})))=
					\end{array}\\[13\jot]
\begin{array}[t]{@{}l@{}}
	\rightI((\PFA,v_{1}))\land\rightI((\PFB,v_{2}))\land\rightI((\PFC,v_{3}))\land\rightI((\PFE,v_{4}))\,\land\\
	\rightI((\PFF,v_{5}))\land\rightI((\PFG,v_{6}))\land\rightI((\PFH,v_{7}))\land\rightI((\PFI,v_{8}))\,\land\\
	\rightI((\PFL,v_{9}))\land\rightI((\PFM,v_{10}))\land\rightI((\PFN,v_{11}))\land\rightI((\PFO,v_{12}))\,\land\\
	\rightI((\PFQone,v_{13}))\land\rightI((\PFQtwo,v_{14}))\land\rightI((\PFQthree,v_{15}))\land\rightI((\PFQfour,v_{16}))\,\land\\
				 	\rightI((\PFPS,v_{17}))\land\rightI((\PFHC,v_{18}))\land\rightI((\PFST,v_{19}))\land\rightI((\PFAD,v_{20}))\,\land\\
					\rightI((\PFLE,v_{21}))\land\rightI((\PFSR,v_{22}))\land\rightI((\PFAW,v_{23}))\land\rightI((\PFPI,v_{24}))\,\land\\
					\rightI((\PFOT,v_{25}))\land\rightI((\PFAP,v_{26}))\land\rightI((\PFTS,v_{27}))\land\rightI((\PFTI,v_{28}))\,,
					\end{array}
				\end{array}$$
\begin{sidewaystable}
\centering
\caption{The translation $\rightI$ of $\PPP\times\{1,2,3,4,5,6,7,8,9,10\}$ to $\LPL(\atoms)$}
\medskip
\resizebox{\textwidth}{!}{
	$\begin{array}{|c|c|c|c|c||c||c|c|c|c|c|}
	\hline 
		\multicolumn{5}{|c||}{\text{Low Range}} & 
		\multirow{2}{4ex}{\centering $\PPP$} & 
		\multicolumn{5}{c|}{\text{High Range}} \\
	\cline{1-5}\cline{7-11}
	1 & 2 & 3 & 4 & 5 & & 6 & 7 & 8 & 9 & 10 \\
	\hline
	\hline
	\F{h}{\mbb} & \F{h}{\mb} & \F{h}{\m} & \F{h}{\m} & \F{h}{\n} & \PFA & 
		\F{h}{\n} & \F{h}{\p} & \F{h}{\p} & \F{h}{\pb} & \F{h}{\pbb}\\
	\hline
	\F{k}{\pbb}\land\F{p}{\mbb} & \F{k}{\pb}\land\F{p}{\mb} & \F{k}{\p}\land\F{p}{\m} & \F{k}{\p}\land\F{p}{\m} & \F{k}{\n}\land\F{p}{\n} & \PFB & 
		\F{k}{\n}\land\F{p}{\n} & \F{k}{\m}\land\F{p}{\p} & \F{k}{\m}\land\F{p}{\p} & \F{k}{\mb}\land\F{p}{\pb} & \F{k}{\mbb}\land\F{p}{\pbb}\\		
	\hline
	\F{d}{\pbb} & \F{d}{\pb} & \F{d}{\p} & \F{d}{\p} & \F{d}{\n} & \PFC & 
		\F{d}{\n} & \F{d}{\m} & \F{d}{\m} & \F{d}{\mb} & \F{d}{\mbb}\\
	\hline
	\F{s}{\mbb} & \F{s}{\mb} & \F{s}{\m} & \F{s}{\m} & \F{s}{\n} & \PFE & 
		\F{s}{\n} & \F{s}{\p} & \F{s}{\p} & \F{s}{\pb} & \F{s}{\pbb}\\
	\hline
	\F{k}{\mbb} & \F{k}{\mb} & \F{k}{\m} & \F{k}{\m} & \F{k}{\n} & \PFF & 
		\F{k}{\n} & \F{k}{\p} & \F{k}{\p} & \F{k}{\pb} & \F{k}{\pbb}\\
	\hline
	\F{e}{\mbb}\land\F{hy}{\pbb}\land\F{k}{\pbb} & \F{e}{\mb}\land\F{hy}{\pb}\land\F{k}{\pb} & \F{e}{\m}\land\F{hy}{\p}\land\F{k}{\p} & \F{e}{\m}\land\F{hy}{\p}\land\F{k}{\p} & \F{e}{\n}\land\F{hy}{\n}\land\F{k}{\n} & \PFG & 
		\F{e}{\n}\land\F{hy}{\n}\land\F{k}{\n} & \F{e}{\p}\land\F{hy}{\m}\land\F{k}{\m} & \F{e}{\p}\land\F{hy}{\m}\land\F{k}{\m} & \F{e}{\pb}\land\F{hy}{\mb}\land\F{k}{\mb} & \F{e}{\pbb}\land\F{hy}{\mbb}\land\F{k}{\mbb}\\		
	\hline
	\F{hy}{\mbb}\land\F{d}{\mbb} & \F{hy}{\mb}\land\F{d}{\mb} & \F{hy}{\m}\land\F{d}{\m} & \F{hy}{\m}\land\F{d}{\m} & \F{hy}{\n}\land\F{d}{\n} & \PFH & 
		\F{hy}{\n}\land\F{d}{\n} & \F{hy}{\p}\land\F{d}{\p} & \F{hy}{\p}\land\F{d}{\p} & \F{hy}{\pb}\land\F{d}{\pb} & \F{hy}{\pbb}\land\F{d}{\pbb}\\		
	\hline
	\F{h}{\mbb}\land\F{hy}{\pbb}\land\F{p}{\pbb} & \F{h}{\mb}\land\F{hy}{\pb}\land\F{p}{\pb} & \F{h}{\m}\land\F{hy}{\p}\land\F{p}{\p} & \F{h}{\m}\land\F{hy}{\p}\land\F{p}{\p} & \F{h}{\n}\land\F{hy}{\n}\land\F{p}{\n} & \PFI & 
		\F{h}{\n}\land\F{hy}{\n}\land\F{p}{\n} & \F{h}{\p}\land\F{hy}{\m}\land\F{p}{\m} & \F{h}{\p}\land\F{hy}{\m}\land\F{p}{\m} & \F{h}{\pb}\land\F{hy}{\mb}\land\F{p}{\mb} & \F{h}{\pbb}\land\F{hy}{\mbb}\land\F{p}{\mbb}\\	
	\hline
	\F{k}{\pbb}\land\F{p}{\pbb} & \F{k}{\pb}\land\F{p}{\pb} & \F{k}{\p}\land\F{p}{\p} & \F{k}{\p}\land\F{p}{\p} & \F{k}{\n}\land\F{p}{\n} & \PFL & 
		\F{k}{\n}\land\F{p}{\n} & \F{k}{\m}\land\F{p}{\m} & \F{k}{\m}\land\F{p}{\m} & \F{k}{\mb}\land\F{p}{\mb} & \F{k}{\mbb}\land\F{p}{\mbb}\\
	\hline
	\F{p}{\mbb} & \F{p}{\mb} & \F{p}{\m} & \F{p}{\m} & \F{p}{\n} & \PFM & 
		\F{p}{\n} & \F{p}{\p} & \F{p}{\p} & \F{p}{\pb} & \F{p}{\pbb}\\
	\hline
	\F{hy}{\pbb} & \F{hy}{\pb} & \F{hy}{\p} & \F{hy}{\p} & \F{hy}{\n} & \PFN & 
		\F{hy}{\n} & \F{hy}{\m} & \F{hy}{\m} & \F{hy}{\mb} & \F{hy}{\mbb}\\
	\hline
	\F{p}{\pbb} & \F{p}{\pb} & \F{p}{\p} & \F{p}{\p} & \F{p}{\n} & \PFO & 
		\F{p}{\n} & \F{p}{\m} & \F{p}{\m} & \F{p}{\mb} & \F{p}{\mbb}\\
	\hline	
	\F{d}{\mbb} & \F{d}{\mb} & \F{d}{\m} & \F{d}{\m} & \F{d}{\n} & \PFQone & 
		\F{d}{\n} & \F{d}{\p} & \F{d}{\p} & \F{d}{\pb} & \F{d}{\pbb}\\
	\hline
	\F{d}{\pbb}\land\F{m}{\pbb} & \F{d}{\pb}\land\F{m}{\pb} & \F{d}{\p}\land\F{m}{\p} & \F{d}{\p}\land\F{m}{\p} & \F{d}{\n}\land\F{m}{\n} & \PFQtwo & 
		\F{d}{\n}\land\F{m}{\n} & \F{d}{\m}\land\F{m}{\m} & \F{d}{\m}\land\F{m}{\m} & \F{d}{\mb}\land\F{m}{\mb} & \F{d}{\mbb}\land\F{m}{\mbb}\\
	\hline
	\F{k}{\pbb} & \F{k}{\pb} & \F{k}{\p} & \F{k}{\p} & \F{k}{\n} & \PFQthree & 
		\F{k}{\n} & \F{k}{\pm} & \F{k}{\pm} & \F{k}{\pmub} & \F{k}{\pmlb}\\
	\hline
	\F{e}{\pbb} & \F{e}{\pb} & \F{e}{\p} & \F{e}{\p} & \F{e}{\n} & \PFQfour & 
		\F{e}{\n} & \F{e}{\m} & \F{e}{\m} & \F{e}{\mb} & \F{e}{\mbb}\\
	\hline
	\hline
	\F{k}{\n}\land\F{p}{\pbb} & \F{k}{\n}\land\F{p}{\pb} & \F{k}{\n}\land\F{p}{\p} & \F{k}{\n}\land\F{p}{\p} & \F{k}{\n}\land\F{p}{\n} & \PFPS & \F{k}{\n}\land\F{p}{\n} & \F{k}{\n}\land\F{p}{\m} & \F{k}{\n}\land\F{p}{\m} & \F{k}{\n}\land\F{p}{\mb} & \F{k}{\n}\land\F{p}{\mbb}\\
	\hline
	\F{hy}{\pbb}\land\F{p}{\pbb} & \F{hy}{\pb}\land\F{p}{\pb} & \F{hy}{\p}\land\F{p}{\p} & \F{hy}{\p}\land\F{p}{\p} & \F{hy}{\n}\land\F{p}{\n} & \PFHC & 
		\F{hy}{\n}\land\F{p}{\n} & \F{hy}{\m}\land\F{p}{\m} & \F{hy}{\m}\land\F{p}{\m} & \F{hy}{\mb}\land\F{p}{\mb} & \F{hy}{\mbb}\land\F{p}{\mbb}\\
	\hline
	\F{s}{\pbb}\land\F{k}{\pbb} & \F{s}{\pb}\land\F{k}{\pb} & \F{s}{\p}\land\F{k}{\p} & \F{s}{\p}\land\F{k}{\p} & \F{s}{\n}\land\F{k}{\n} & \PFST & 
		\F{s}{\n}\land\F{k}{\n} & \F{s}{\m}\land\F{k}{\m} & \F{s}{\m}\land\F{k}{\m} & \F{s}{\mb}\land\F{k}{\mb} & \F{s}{\mbb}\land\F{k}{\mbb}\\
	\hline
	\F{p}{\pbb}\land\F{d}{\mbb} & \F{p}{\pb}\land\F{d}{\mb} & \F{p}{\p}\land\F{d}{\m} & \F{p}{\p}\land\F{d}{\m} & \F{p}{\n}\land\F{d}{\n} & \PFAD & 
		\F{p}{\n}\land\F{d}{\n} & \F{p}{\m}\land\F{d}{\p} & \F{p}{\m}\land\F{d}{\p} & \F{p}{\mb}\land\F{d}{\pb} & \F{p}{\mbb}\land\F{d}{\pbb}\\
	\hline
	\F{s}{\pbbb}\lor\F{s}{\mbbb} & \F{s}{\pbbb}\lor\F{s}{\mbbb} & \F{s}{\pbb}\lor\F{s}{\mbb} & \F{s}{\pbb}\lor\F{s}{\mbb} & \F{s}{\pb}\lor\F{s}{\mb} & \PFLE & \F{s}{\pb}\lor\F{s}{\mb} & \F{s}{\p}\lor\F{s}{\m} & \F{s}{\p}\lor\F{s}{\m} & \F{s}{\n} & \F{s}{\n}\\
	\hline
	\F{s}{\pbb}\land\F{k}{\pbb}\land\F{p}{\mbb} & \F{s}{\pb}\land\F{k}{\pb}\land\F{p}{\mb} & \F{s}{\p}\land\F{k}{\p}\land\F{p}{\m} & \F{s}{\p}\land\F{k}{\p}\land\F{p}{\m} & \F{s}{\n}\land\F{k}{\n}\land\F{p}{\n} & \PFSR & 
		\F{s}{\n}\land\F{k}{\n}\land\F{p}{\n} & \F{s}{\m}\land\F{k}{\m}\land\F{p}{\p} & \F{s}{\m}\land\F{k}{\m}\land\F{p}{\p} & \F{s}{\mb}\land\F{k}{\mb}\land\F{p}{\pb} & \F{s}{\mbb}\land\F{k}{\mbb}\land\F{p}{\pbb}\\
	\hline
	\F{d}{\pbb}\land\F{m}{\pbb} & \F{d}{\pb}\land\F{m}{\pb} & \F{d}{\p}\land\F{m}{\p} & \F{d}{\p}\land\F{m}{\p} & \F{d}{\n}\land\F{m}{\n} & \PFAW & 
		\F{d}{\n}\land\F{m}{\n} & \F{d}{\m}\land\F{m}{\m} & \F{d}{\m}\land\F{m}{\m} & \F{d}{\mb}\land\F{m}{\mb} & \F{d}{\mbb}\land\F{m}{\mbb}\\
	\hline
	(\F{k}{\pmlb}\lor\F{k}{\pmub})\land\F{p}{\n} & (\F{k}{\pmlb}\lor\F{k}{\pmub})\land\F{p}{\n} & \F{k}{\pm}\land\F{p}{\n} & \F{k}{\pm}\land\F{p}{\n} & \F{k}{\n}\land\F{p}{\n} & \PFPI & 
		\F{k}{\n}\land\F{p}{\n} & \F{k}{\n}\land\F{p}{\pm} & \F{k}{\n}\land\F{p}{\pm} & \F{k}{\n}\land(\F{p}{\pmlb}\lor\F{p}{\pmub}) & \F{k}{\n}\land(\F{p}{\pmlb}\lor\F{p}{\pmub})\\
	\hline
	\F{k}{\n}\land\F{p}{\mbb} & \F{k}{\n}\land\F{p}{\mb} & \F{k}{\n}\land\F{p}{\m} & \F{k}{\n}\land\F{p}{\m} & \F{k}{\n}\land\F{p}{\n} & \PFOT & 
		\F{k}{\n}\land\F{p}{\n} & \F{k}{\pm}\land\F{p}{\p} & \F{k}{\pm}\land\F{p}{\p} & (\F{k}{\pmlb}\lor\F{k}{\pmub})\land\F{p}{\pb} & (\F{k}{\pmlb}\lor\F{k}{\pmub})\land\F{p}{\pbb}\\
	\hline
	\F{k}{\pbb}\land\F{p}{\n} & \F{k}{\pb}\land\F{p}{\n} & \F{k}{\p}\land\F{p}{\n} & \F{k}{\p}\land\F{p}{\n} & \F{k}{\n}\land\F{p}{\n} & \PFAP & 
		\F{k}{\n}\land\F{p}{\n} & \F{k}{\m}\land\F{p}{\pm} & \F{k}{\m}\land\F{p}{\pm} & \F{k}{\mb}\land(\F{p}{\pmlb}\lor\F{p}{\pmub}) & \F{k}{\mbb}\land(\F{p}{\pmlb}\lor\F{p}{\pmub})\\
	\hline
	\F{e}{\pbb}\land\F{d}{\mbb} & \F{e}{\pb}\land\F{d}{\mb} & \F{e}{\p}\land\F{d}{\m} & \F{e}{\p}\land\F{d}{\m} & \F{e}{\n}\land\F{d}{\n} & \PFTS & 
		\F{e}{\n}\land\F{d}{\n} & \F{e}{\m}\land\F{d}{\p} & \F{e}{\m}\land\F{d}{\p} & \F{e}{\mb}\land\F{d}{\pb} & \F{e}{\mbb}\land\F{d}{\pbb}\\
	\hline
	\F{hy}{\mbb}\land\F{p}{\mbb}\land\F{d}{\mbb} & \F{hy}{\mb}\land\F{p}{\mb}\land\F{d}{\mb} & \F{hy}{\m}\land\F{p}{\m}\land\F{d}{\m} & \F{hy}{\m}\land\F{p}{\m}\land\F{d}{\m} & \F{hy}{\n}\land\F{p}{\n}\land\F{d}{\n} & \PFTI & 
		\F{hy}{\n}\land\F{p}{\n}\land\F{d}{\n} & \F{hy}{\p}\land\F{p}{\p}\land\F{d}{\p} & \F{hy}{\p}\land\F{p}{\p}\land\F{d}{\p} & \F{hy}{\pb}\land\F{p}{\pb}\land\F{d}{\pb} & \F{hy}{\pbb}\land\F{p}{\pbb}\land\F{d}{\pbb}\\
	\hline
\end{array}$}
\label{table:PPPtoLPL}
\end{sidewaystable}
			\item the function space $(2^{\PPPts}\to\LPL(\atoms))$ such that for every $F\in2^{\PPPts}$, 
						$$\rightI(F) = \bigwedge\{\,\rightI(f) \mid f\in F\,\}\,;$$
		\end{itemize}
			\item $\leftI$ be defined in the function space $(\SPPts\to\LPL(\atoms))$ such that 
					$$\begin{array}{@{}r@{}}
						 \leftI(((\F{h}{,s_{1}}), (\F{s}{,s_{2}}), (\F{e}{,s_{3}}), (\F{hy}{,s_{4}}), 
			(\F{k}{,s_{5}}), (\F{p}{,s_{6}}), (\F{d}{,s_{7}}), (\F{m}{,s_{8}})))=\\
						\F{h}{s_{1}}\land\F{s}{s_{2}}\land\F{e}{s_{3}}\land\F{hy}{s_{4}}\land\F{k}{s_{5}}\land\F{p}{s_{6}}\land\F{d}{s_{7}}\land\F{m}{s_{8}}
					\end{array}$$
					and in the function space $(2^{\SPPts}\to\LPL(\atoms))$ such that for every $P\in2^{\SPPts}$,
						$$ \leftI(P) = \bigvee\{\, \leftI(p) \mid p\in P\,\}\,.$$
		\end{itemize}
	Then, the mapping  
		\begin{itemize}
			\item $\rightG{}:\PPPps\to\SPPps$ defined such that for every $F\in2^{\PPPts}$, 
					$$\rightG{F} = \{\,p\in\SPPts \mid \leftI(p)\Rightarrow\rightI(F)\,\}$$
					is the so-called \emph{right polarity} and
			\item $\leftG{}:\SPPps\to\PPPps$ defined such that for every $P\in2^{\SPPts}$, 
					$$\leftG{P} = \{\,f\in\PPPts \mid \leftI(P)\Rightarrow\rightI(f)\,\}$$
					is the so-called \emph{left polarity} of the ordered pair $(\,\rightG{},\leftG{}\,)$.
		\end{itemize}		
\end{definition}
\noindent 
Spelled out, 
	(1) the result of 
			applying the mapping $\rightI$ to 
				a set $F$ of PPPs $f$ as defined in Definition~\ref{definition:MappingsAndMorphisms} is 
					the conjunction of the results of 
						applying $\rightI$ to 
							each one of these $f$, which in turn 
								is the conjunction of the results of
									applying $\rightI$ to 
										each one of the factor-value pairs in $f$ as 
											defined in Table~\ref{table:PPPtoLPL};
	(2) the result of 
			applying the mapping $\leftI$ to 
				a set $P$ of SPPs $p$ as defined in Definition~\ref{definition:MappingsAndMorphisms} is 
					the disjunction of the results of 
						applying $\leftI$ to 
							each one of these $p$, which 
								simply is the conjunction of 
									all signed factors in $p$ taken each one as an atomic proposition;
	(3)	the result of 
			applying the mapping $\rightG{}$ to
				a set $F$ of PPPs is 
					the set of all those SPPs $p$ whose 
						mapping under $\leftI$ implies the mapping of $F$ under $\rightI$;
	(4) the result of 
			applying the mapping $\leftG{}$ to
				a set $P$ of SPPs is 
					the set of all those PPPs $f$ whose 
						mapping under $\rightI$ is implied by the mapping of $P$ under $\leftI$.	
Thus from a computer science perspective \cite[Section~7.35]{DaveyPriestley}, 
	PPPs are specifications of SPPs and 
	SPPs are implementations or refinements of PPPs.
The Galois-connection then connects correct implementations to their respective specification by 
	stipulating that a correct implementation imply its specification.
By convention,
	$\bigwedge\emptyset=\top$ and $\bigvee\emptyset=\bot$\,, that is,
		the conjunction over the empty set $\emptyset$ is tautological truth $\top$\,, and 
		the disjunction over $\emptyset$ is tautological falsehood $\bot$\,, respectively.

Note that an example of an SPP that 
	maps to the empty set under $\leftG{}$ happens to be the Szondi norm profile mentioned before, because 
		its mapping under $\leftI$ 
	$$\begin{array}{r}
		\leftI(((\F{h}{,\p}), (\F{s}{,\p}), (\F{e}{,\m}), (\F{hy}{,\m}), 
				 (\F{k}{,\m}), (\F{p}{,\m}), (\F{d}{,\p}), (\F{m}{,\p})))=\\
				 	\F{h}{\p}\land\F{s}{\p}\land\F{e}{\m}\land\F{hy}{\m}\land\F{k}{\m}\land\F{p}{\m}\land\F{d}{\p}\land\F{m}{\p}\,,
	\end{array}$$
does not meet our translation of Cattell's personality trait 
	$\PFB$, $\PFG$, $\PFH$, $\PFM$, $\PFQthree$, $\PFPS$, $\PFST$, $\PFLE$, $\PFSR$, $\PFPI$, $\PFOT$, $\PFAP$, $\PFTS$, nor $\PFTI$, as can seen by inspecting Table~\ref{table:PPPtoLPL}.

As can also be seen in Table~\ref{table:PPPtoLPL},
	our interpretation of Cattell's scale is mostly the following: 
		Cattell's value $1$ becomes Szondi's signature $\mbb$,
		$2$ becomes $\mb$,
		$3$ and $4$ become $\m$,
		$5$ and $6$ become $\n$,
		$7$ and $8$ become $\p$,
		$9$ becomes $\pb$, and 
		$10$ becomes $\pbb$.
This corresponds to 
	how Szondi accounts for 
		the corresponding number of 
			portrait choices of the same kind in his test \cite{Szondi:ETD:Band1}:
				the low range $1$--$5$ corresponds to the numbers $1$--$5$ of antipathy choices (portrait dislikes), respectively, and
				the high range $6$--$10$ to the numbers $1$--$5$ of sympathy choices (portrait likes), respectively.
Of course, our readers may experiment with their own interpretation and accounting. 
For example, 
	they might want to take into account also 
		Szondi's signatures $\mbbb$ and $\pbbb$ for pathologically strong, unambiguous negative and positive choices, respectively, and
			adapt the scale accordingly.
Szondi's signatures $\m$ and $\p$ account for normally strong, unambiguous negative and positive choices, respectively.
Szondi's test also allows for ambiguous sets of (portrait) choices 
	(noted---``signed'' in Szondi's terminology---as $\pm$, $\pmlb$, and $\pmub$).
This ambiguity turns out to be also useful in our translation in Table~\ref{table:PPPtoLPL}.
Observe that 
	(1) in the translation of the low-high range opposition, 
			we have made use of 
				signature opposition (polarity, e.g., $\F{h}{\m}$ versus $\F{h}{\p}$);
	(2) abnormal personality traits translate all into psychological syndromes, that is,
		conjunctions of signed factors; and 
	(3) any conjunctive low-range translation is 
			the conjunction of the opposed factors of the corresponding high range translation.
This last observation makes PPPs appear quite rigid, but 
	is justified by the (natural-language) definition---``descriptors'' in Cattell's terminology---of \mbox{Cattell's} personality traits \cite[Table~7.1]{16PFinHB}, which
		we recall by annotating them with Szondi's signed factors (Cattell's commas correspond to conjunctions here):
	\begin{enumerate}
		\item Reserved [\F{h}{\m}], Impersonal [\F{h}{\m}], Distant [\F{h}{\m}]---Warmth 
			(\PFA)---Warm-\linebreak hearted [\F{h}{\p}], Caring [\F{h}{\p}], Attentive To Others [\F{h}{\p}];
		\item Concrete [\F{k}{\p}, having, matter], Lower Mental Capacity [\F{p}{\m}, psychological projection, subjectivity]---Reasoning 
			(\PFB)---Abstract [\F{p}{\p}, being, ideas], Bright [\F{k}{\m}, \F{p}{\p}], Fast-Learner [\F{p}{\p}, intuition];
		\item Reactive [\F{s}{\m}, \F{d}{\p}], Affected By Feelings [\F{d}{\p}, depression]---Emotional Stability 
			(\PFC)---Emotionally Stable [\F{d}{\m}], Adaptive Mature [\F{d}{\pm}];
				
		\item Deferential [\F{s}{\m}], Cooperative [\F{s}{\m}], Avoids Conflict [\F{s}{\m}]---Dominance\linebreak  
			(\PFE)---Dominant [\F{s}{\p}], Forceful [\F{s}{\p}], Assertive [\F{s}{\p}];
		\item Serious [\F{k}{\m}], Restrained [\F{k}{\m}], Careful [\F{k}{\m}]---Liveliness 
			(\PFF)---Enthusiastic [\F{k}{\p}], Animated [\F{k}{\p}], Spontaneous [\F{k}{\p}];
		\item Expedient [\F{e}{\m}, \F{hy}{\p}, \F{k}{\p}], Nonconforming [\F{e}{\m}, \F{hy}{\p}]---Rule-Consciousness 
			(\PFG)---Rule-Conscious [\F{e}{\p}, \F{hy}{\m}, \F{k}{\m}], Dutiful [\F{e}{\p}];
		\item Shy [\F{hy}{\m}], Timid [\F{hy}{\m}], Threat-Sensitive [\F{d}{\m}]---Social Boldness 
			(\PFH)---Socially Bold [\F{hy}{\p}], Venturesome [\F{d}{\p}], Thick-Skinned [\F{h}{\n}];
		\item Tough [\F{h}{\n}, \F{hy}{\p}, \F{p}{\p}], Objective [\F{p}{\p}], Unsentimental [\F{h}{\m}]---Sensitivity
			(\PFI)---Sensitive [\F{h}{\p}, \F{hy}{\m}, \F{p}{\m}], Aesthetic [\F{h}{\p}], Tender-Minded [\F{h}{\p}, \F{p}{\m}];
		\item Trusting [\F{p}{\p}, \F{m}{\p}], Unsuspecting [\F{p}{\p}], Accepting [\F{k}{\p}]---Vigilance 
			(\PFL)---Vigilant [\F{p}{\m}], Suspicious [\F{p}{\m}], Skeptical [\F{k}{\m}], Wary [\F{p}{\m}];
		\item Practical [\F{p}{\m}], Grounded [\F{p}{\m}], Down-To-Earth [\F{p}{\m}]---Abstractedness\linebreak  
			(\PFM)---Abstracted [\F{p}{\p}], Imaginative [\F{p}{\p}], Idea-Oriented [\F{p}{\p}]; 
		\item Forthright [\F{hy}{\p}], Genuine [\F{hy}{\p}], Artless [\F{hy}{\p}]---Privateness 
			(\PFN)---Private [\F{hy}{\m}], Discreet [\F{hy}{\m}], Non-Disclosing [\F{hy}{\m}];
		\item Self-Assured [\F{p}{\p}], Unworried [\F{p}{\p}], Complacent [\F{p}{\p}]---Apprehension\linebreak 
			(\PFO)---Apprehensive [\F{p}{\m}], Self-Doubting [\F{p}{\m}], Worried [\F{p}{\m}];
		\item Traditional [\F{d}{\m}], Attached To Familiar [\F{d}{\m}]---Openness to Change 
			(\PFQone)---Open To Change [\F{d}{\p}], Experimenting [\F{d}{\p}];
		\item Group-Oriented [\F{d}{\p}, \F{m}{\p}], Affiliative [\F{d}{\p}, \F{m}{\p}]---Self-Reliance 
			(\PFQtwo)---Self-Reliant [\F{d}{\m}, \F{m}{\m}], Solitary [\F{d}{\m}, \F{m}{\m}], Individualistic [\F{d}{\m}, \F{d}{\m}];
		\item Tolerates Disorder [\F{k}{\n}], Unexacting [\F{k}{\n}], Flexible [\F{k}{\n}]---Perfectionism\linebreak 
			(\PFQthree)---Perfectionistic [\F{k}{\pm}], Organized [\F{k}{\m}], Self-Disciplined [\F{k}{\pm}];
		\item Relaxed [\F{e}{\p}], Placid [\F{e}{\p}], Patient [\F{e}{\p}]---Tension
			(\PFQfour)---Tense [\F{e}{\m}], High Energy [\F{e}{\m}], Driven [\F{e}{\m}].
	\end{enumerate}
	
Cattell's global personality factors  
	(Cattell's ``Big Five''),    
	defined as groups of 16PF primary traits \cite[Table~7.2]{16PFinHB}, 
		can then simply be translated as 
			disjunctions of the translations of 
				the corresponding primary traits.
That is, for every value $v\in\{1,2,3,4,5,6,7,8,9,10\}:$
	\begin{eqnarray*}
		\text{Extraversion $v$} & = & 
			\begin{array}[t]{@{}l@{}}
				\rightI((\PFA,v))\lor\rightI((\PFF,v))\lor\rightI((\PFH,v))\,\lor\\
				\rightI((\PFN,10-v))\lor\rightI((\PFQtwo,10-v))
			\end{array}\\
		\text{High Anxiety $v$} & = & 
			\rightI((\PFC,v))\lor\rightI((\PFL,v))\lor\rightI((\PFO,v))\lor\rightI((\PFQfour,v))\\
		\text{Tough-Mindedness $v$} & = & 
			\rightI((\PFA,10-v))\lor\rightI((\PFI,10-v))\lor\rightI((\PFM,v))\lor\rightI((\PFQone,v))\\
		\text{Independence $v$} & = & 
			\rightI((\PFE,v))\lor\rightI((\PFH,v))\lor\rightI((\PFL,10-v))\lor\rightI((\PFQone,v))\\
		\text{Self-Control $v$} & = & 
			\rightI((\PFF,10-v))\lor\rightI((\PFG,v))\lor\rightI((\PFM,10-v))\lor\rightI((\PFQthree,v))
	\end{eqnarray*}

Like in \cite{arXiv:1403.2000v1}, 
	we now prove in two intermediate steps that 
		the pair $(\,\rightG{},\leftG{}\,)$ is indeed a Galois-connection.
The first step is the following announced fact, from which 
	the second step, Lemma~\ref{lemma:Properties}, follows, from which in turn 
		the desired result, Theorem~\ref{theorem:Galois}, then follows---easily.
As announced,
	all that our readers need to check on their own analog of our mapping $\rightI$ is 
		that it has the property stated as Fact~\ref{fact:FactsAboutip}.1.
Their own Galois-connection is then automatically induced.
\begin{fact}[Some facts about $\rightI$ and $\leftI$]\label{fact:FactsAboutip}\ 
	\begin{enumerate}
		\item if $F\subseteq F'$ then $\rightI(F')\Rightarrow\rightI(F)$
		\item if $P\subseteq P'$ then $\leftI(P)\Rightarrow\leftI(P')$
		\item The function $\leftI$ but not the function $\rightI$ is injective, and
				neither is surjective.
	\end{enumerate}
\end{fact}
\begin{proof}
	By inspection of Definition~\ref{definition:MappingsAndMorphisms} and Table~\ref{table:PPPtoLPL}.
\end{proof}
\noindent
Like in \cite{arXiv:1403.2000v1}, 
	we need 
		Fact~\ref{fact:FactsAboutip}.1 and \ref{fact:FactsAboutip}.2 but 
		not Fact~\ref{fact:FactsAboutip}.3 in the following development.
Therefor, note the two macro-definitions 
	$\rightleftG{}:=\rightG{}\circ\leftG{}$ and 
	$\leftrightG{}:=\leftG{}\circ\rightG{}$ with 
		$\circ$ being function composition, as usual (from right to left, as usual too).
\begin{lemma}[Some useful properties of $\rightG{}$ and $\leftG{}$]\label{lemma:Properties}\ 
	\begin{enumerate}
		\item if $F\subseteq F'$ then $\rightG{F'}\subseteq\rightG{F}$\quad(\;$\rightG{}$ is antitone)
		\item if $P\subseteq P'$ then $\leftG{P'}\subseteq\leftG{P}$\quad(\,$\leftG{}$ is antitone)
		\item $P\subseteq\rightG{(\leftG{P})}$\quad(\;$\rightleftG{}$ is inflationary)
		\item $F\subseteq\leftG{(\rightG{F})}$\quad(\,$\leftrightG{}$ is inflationary)
	\end{enumerate}
\end{lemma}
\begin{proof}
	Like in \cite{arXiv:1403.2000v1}.
\end{proof}
\noindent
We are ready for making the final step.
\begin{theorem}[The Galois-connection property of $(\,\rightG{},\leftG{}\,)$]\label{theorem:Galois}
	The ordered pair $(\,\rightG{},\leftG{}\,)$ is an \emph{antitone} or \emph{order-reversing Galois-connection} between $\PPPps$ and $\SPPps$.
	That is, 
		for every $F\in2^{\PPPts}$ and $P\in2^{\SPPts}$,
			$$\text{$P\subseteq\rightG{F}$ if and only if $F\subseteq\leftG{P}$.}$$
\end{theorem}
\begin{proof}
	Like in \cite{arXiv:1403.2000v1}.
\end{proof}
\noindent
Thus from a computer science perspective \cite[Section~7.35]{DaveyPriestley}, 
	smaller (larger) sets of PPPs and thus less (more) restrictive specifications correspond to 
	larger (smaller) sets of SPPs and thus more (fewer) possible implementations.

Note that Galois-connections are 
	connected to \emph{residuated mappings} \cite{LatticesAndOrderedAlgebraicStructures}.
Further, 
	natural notions of equivalence on $\PPPps$ and $\SPPps$ are given by 
		the \emph{kernels} of $\rightG{}$ and $\leftG{}$, respectively, which are, by definition:
$$\begin{array}{rcl}
	F\equiv F' &\text{if and only if}& \rightG{F}=\rightG{F'}\;;\\[\jot]
	P\equiv P' &\text{if and only if}& \leftG{P}=\leftG{P'}\,.
\end{array}$$

\begin{proposition}[The computability of $(\,\rightG{},\leftG{}\,)$]\ 
	\begin{enumerate}
		\item Given $F\in2^{\PPPts}$, $\rightG{F}$ is computable.
		\item Given $P\in2^{\SPPts}$, $\leftG{P}$ is computable.
	\end{enumerate}
\end{proposition}
\begin{proof}
	Similar to \cite{arXiv:1403.2000v1}, but
		with the difference that 
			the Galois-connection there is efficiently computable,
				whereas the one here is only so for small sets $F$ and $P$
					(which in practice usually are singleton sets of only one personality profile).
\end{proof}

\section{Conclusion}
We have proposed a computable Galois-connection between 
	PsychEval Personality Profiles (including the 16PF Personality Profiles) and 
	Szondi's personality profiles,  
		as promised in the abstract and 
		as a further illustration of 
			our simple methodology introduced in \cite{arXiv:1403.2000v1} 
				for generating such Galois-connections.

\paragraph{Acknowledgements} 
The \LaTeX-package TikZ was helpful for graph drawing.

\bibliographystyle{plain}

\end{document}